\documentclass[twoside]{dis09}
\usepackage[latin1]{inputenc}
\usepackage[dvips]{graphicx,epsfig,color}
\usepackage{wrapfig,rotating}
\usepackage{amssymb,amsmath,array}

\begin{document}
\title{Longitudinal target polarization dependence of $\bar \Lambda$
polarization and polarized strangeness PDFs}

\author{Aram Kotzinian
%
\thanks{Talk~\cite{url} delivered by at the XVII International Workshop on Deep-Inelastic Scattering and
Related Subjects, DIS 2009, 26-30 April 2009, Madrid, Spain.}
%
\vspace{.3cm}\\
%
CEA-Saclay, IRFU/Service de Physique Nucléaire, 91191 Gif-sur-Yvette, France\\
Yerevan Physics Institute, Armenia and JINR, Russia
}

\maketitle

\begin{abstract}
The longitudinal polarization of $\bar \Lambda$ produced in the current fragmentation region of polarized lepton DIS off polarized and unpolarized target is described both in the simple formalism with factorized fragmentation functions as well as within intrinsic strangeness model using string fragmentation implemented into event generator LEPTO.

It is demonstrated that the the measurement of $\bar \Lambda$ polarization and its dependence on the target polarization can serve as a filter for (un)polarized anti-strangeness distribution function of nucleon.

\end{abstract}

\section{Introduction}

According to QED in the hard charged longitudinally polarized lepton--unpolarized quark scattering $l+q \to l^{\prime}+q^{\prime}$ lepton transfers its polarization to final quark. Within the simple parton model description of SIDIS the hadron production in the current fragmentation region is described by independent fragmentation of scattered quark. Then the (anti)quark polarization can be transferred to produced $\bar \Lambda$. For unpolarized target the final quark polarization, $P_{q^{\prime}}$, is related to the beam polarization, $P_B$, by
\begin{equation}\label{qpol}
P_{q^{\prime}}=D(y)P_B,
\end{equation}
where $D(y)=\frac{y(1-y)}{1+(1-y)^2}$ is the depolarization factor.
The resulting $\bar \Lambda$ polarization in this simple approach is given by
\begin{equation}\label{alpol}
P^{\bar \Lambda}=P^{\bar \Lambda}_{P_B,0}=D(y)P_B\frac
{\sum_q\,e_q^2 q(x) \Delta D_q^{\bar \Lambda}}
{\sum_q\,e_q^2 q(x) D_q^{\bar \Lambda}},
\end{equation}
where subscript in $P^{\bar \Lambda}_{P_B,P_T}$ denotes the beam ($P_B$) and target ($P_T$) longitudinal polarization.

Here we are considering only $\bar \Lambda$ longitudinal polarization. The description of $\Lambda$ polarization is more involved since the target remnant polarized diquark is also participating in polarization transfer. More details and discussion can be found in \cite{ekk}-\cite{ekns}.

Within the static SU(6) model the whole spin of $\bar \Lambda$ is carried by the $\bar s$-quark therefore the polarized $\bar s$ fragmentation function $\Delta D_{\bar s}^{\bar \Lambda}=D_{\bar s}^{\bar \Lambda}$ and other quarks does not participate in spin transfer. Then for longitudinal (along the virtual photon direction) spin transfer, $S_X^{\bar \Lambda}$ we have
\begin{equation}\label{su6st}
S_X^{\bar \Lambda} \doteq \frac{P^{\bar \Lambda}}{D(y)P_B} =
\frac{e_{\bar s}^2 {\bar s}(x) D_{\bar s}^{\bar \Lambda}}
{\sum_q\,e_q^2 q(x) D_q^{\bar \Lambda}},
\end{equation}
which is nothing else than relative contribution (purity) of $\bar s$-quarks in ${\bar \Lambda}$ production.

More general approach was considered in \cite{ekk} -- \cite{ekns}. In contrast to the simple independent fragmentation approach in these articles the spin transfer from intrinsic polarized strangeness as well as from target remnant polarized diquark was also taken into account. The calculations were performed using the LUND string fragmentation model for hadronization of state originated after hard scattering. Two different assumptions were considered
\begin{itemize}
  \item Model {\bf A}: the spin transfer to $\bar \Lambda$ or to intermediate heavy hyperons is nonzero only for first rank hyperons,
  \item Model {\bf B}: the spin transfer to $\bar \Lambda$ or to intermediate heavy hyperons is nonzero for all ranks hyperons.
\end{itemize}
In the following we present the results of our calculations with Model {\bf B} using the intrinsic strangeness correlation coefficients, $c_{sq}$, obtained in \cite{ekn} and assuming $c_{{\bar s}q}=c_{sq}$.

\begin{figure}[h]
 \hspace{-0.5cm}
  \begin{minipage}[t]{.45\textwidth}
    \begin{center}
      \epsfig{file=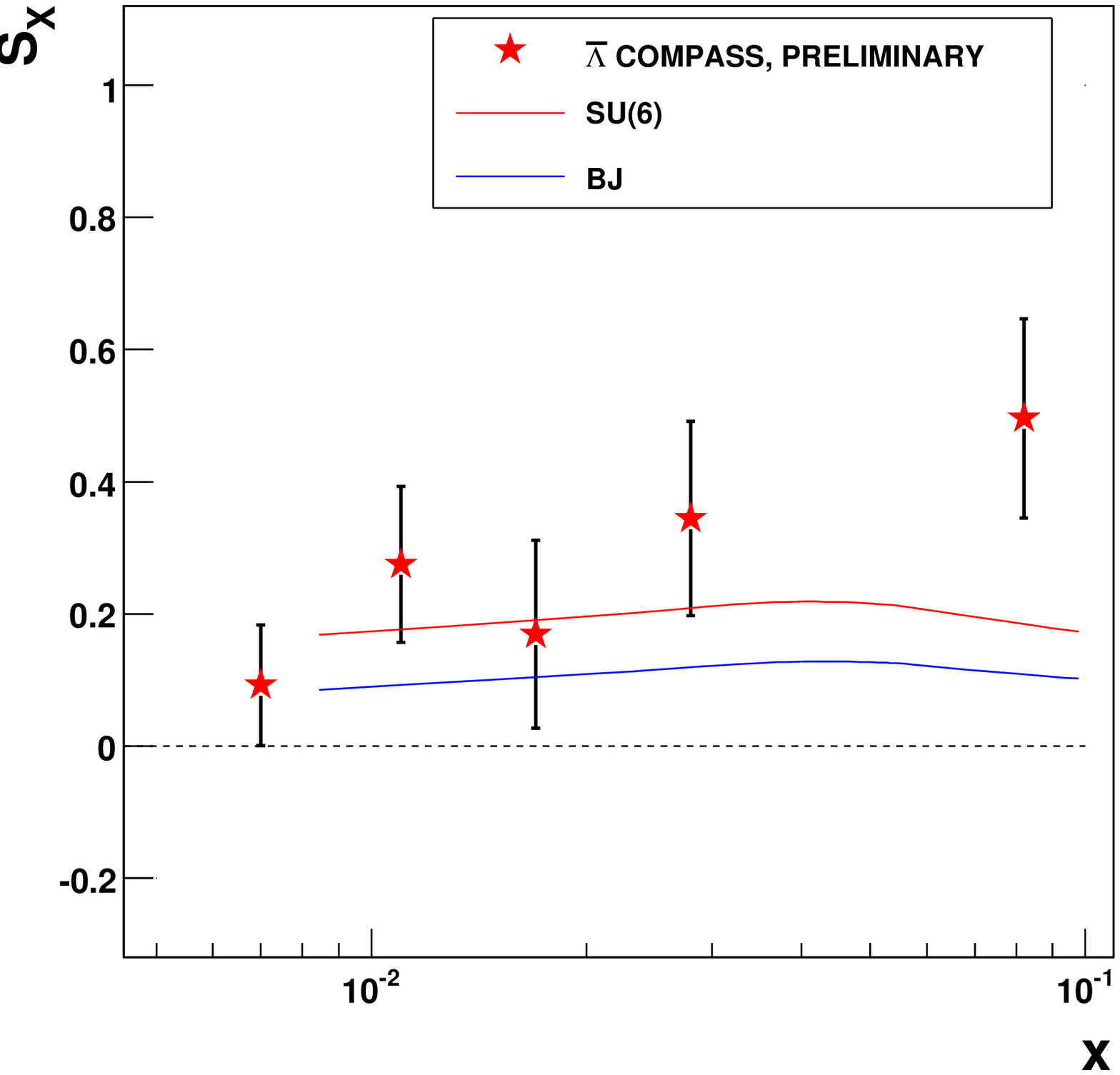, scale=0.395}
    \end{center}
  \end{minipage}
  \hspace{0.8cm}
  \begin{minipage}[t]{.45\textwidth}
    \begin{center}
      \epsfig{file=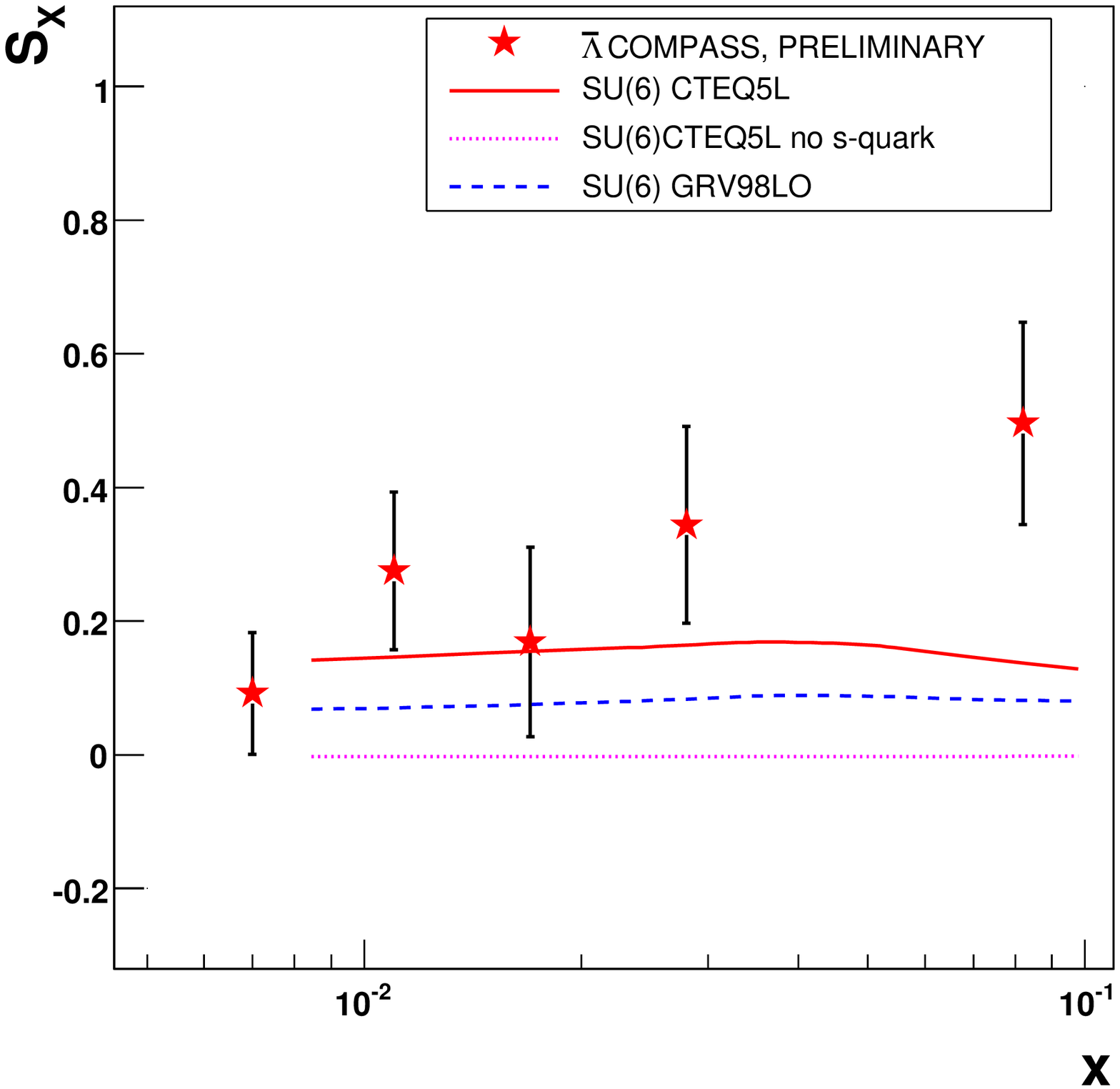, scale=0.4}
    \end{center}
  \end{minipage}
      \caption{Left panel: Spin transfer to $\bar \Lambda$ in SU(6) and BJ models calculated with GRV98LO PDFs. Right panel: Spin transfer to $\bar \Lambda$ in SU(6) model for GRV98LO and CTEQ5L PDFs.}
      \label{fig:alpol}
\end{figure}

\begin{figure}[h]
    \begin{center}
      \epsfig{file=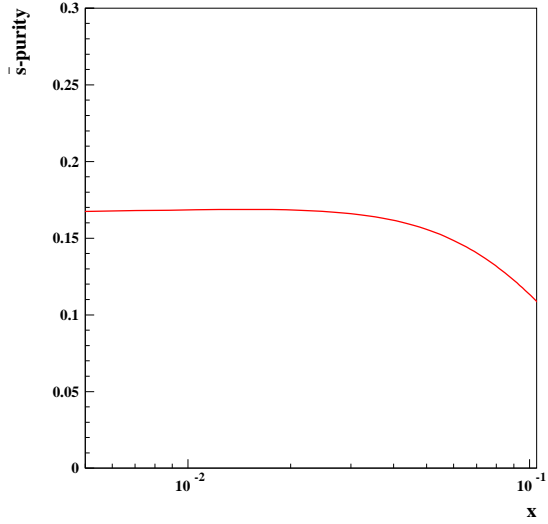, scale=0.395}
    \end{center}
      \caption{The relative contribution of $\bar s$ quarks in $\bar \Lambda$ production for COMPASS cuts. CTEQ5L PDFs were used.}
      \label{fig:sb-pur}
\end{figure}

In Fig. \ref{fig:alpol} the results of full calculations \cite{ekns} (using the cuts corresponding to COMPASS data) are presented together with recent data from COMPASS experiment \cite{sap}, \cite{Alekseev:2009tv}. On the left panel the upper curve corresponds to SU(6) model for spin transfer from quark to hyperons and lower curve -- to Burkardt-Jaffe mechanism \cite{BJ}. COMPASS data indicate that the SU(6) model better describes the spin transfer. On the right panel the results with SU(6) spin transfer and different choices of PDFs are presented: upper curve corresponds to CTEQ5L, middle dashed curve -- to GRV98LO and lower curve to CTEQ5L with strangeness distribution putted equal to zero. First of all one see that when strangeness PDF is switched off the predicted spin transfer becomes negligible.  One can see also that the better description is achieved with CTEQ5L PDFs where the strangeness content is higher than in GRV98LO. In is interesting to note that the upper curves on in Fig. \ref{fig:alpol} are practically repeating the relative contribution of $\bar s$-quarks (purity) calculated according to Eq. (\ref{su6st}) using the LEPTO MC (see Fig. \ref{fig:sb-pur}).

 It is well known that the production rate (multiplicity) of $\bar \Lambda$ is dominated by the scattering off $u$-quarks in nucleon. In contrast, as one can see from Eg. \ref{alpol}  and complete calculations in Fig. \ref{fig:alpol} and $\bar s$-purity in Fig. \ref{fig:sb-pur} the $\bar \Lambda$ polarization is defined mainly by ${\bar s}$ distribution. This, the measurements of $\bar \Lambda$ polarization in the current fragmentation region of SIDIS can serve as a filter to study the strangeness distribution in nucleon.

\section{Target polarization dependence}

In the case of longitudinally polarized target there are two sources of the final quark polarization: the spin transfer from polarized lepton and from polarized target. In this case the independent fragmentation approach predicts
\begin{equation}\label{alpolp}
P^{\bar \Lambda}_{P_B,P_T}=\frac
{\sum_q\,e_q^2 \left [D(y)P_B q(x)-fP_T \Delta q \right ]\Delta D_q^{\bar \Lambda}}
{\sum_q\,e_q^2 \left [q(x)- D(y)P_BfP_T \Delta q \right ]D_q^{\bar \Lambda}},
\end{equation}
where $f$ stands for the target polarization dilution factor. For COMPASS target the contribution of second term in square brackets in the denominator of the above equation is less than 7\% for GRSV \cite{grsv} or recent DSSV \cite{dssv} polarized PDFs. Neglecting this term and again assuming the simple SU(6) model for spin transfer one gets
\begin{equation}\label{dalpol}
\Delta P^{\bar \Lambda}=P^{\bar \Lambda}_{P_B,-P_T}-P^{\bar \Lambda}_{P_B,+P_T}=2fP_T
\frac{\Delta{\bar s}(x)}{{\bar s}(x)}
\frac{e_{{\bar s}}^2 {\bar s}(x) \Delta D_{\bar s}^{\bar \Lambda}}
{\sum_q\,e_q^2 q(x) D_q^{\bar \Lambda}},
\end{equation}
where $+P_T$ ($-P_T$) corresponds to target polarized along (opposite) to beam direction.

Thus, with above approximations the contributions of spin transfer from lepton for opposite target polarizations is canceling and we have direct access to the anti-strangeness polarization by measuring $\Delta P^{\bar \Lambda}$. Comparing Eqs. (\ref{dalpol}) and (\ref{alpol}) we obtain the following simple expression for the anti-strangeness polarization:
\begin{equation}\label{ds-s}
\frac{\Delta{\bar s}(x)}{{\bar s}(x)} \simeq
\frac{\langle D(y) \rangle P_B}{2fP_T}
\frac{\Delta P^{\bar \Lambda}}{P^{\bar \Lambda}}.
\end{equation}

\begin{figure}[h]
 \hspace{-0.5cm}
  \begin{minipage}[t]{.45\textwidth}
    \begin{center}
      \epsfig{file=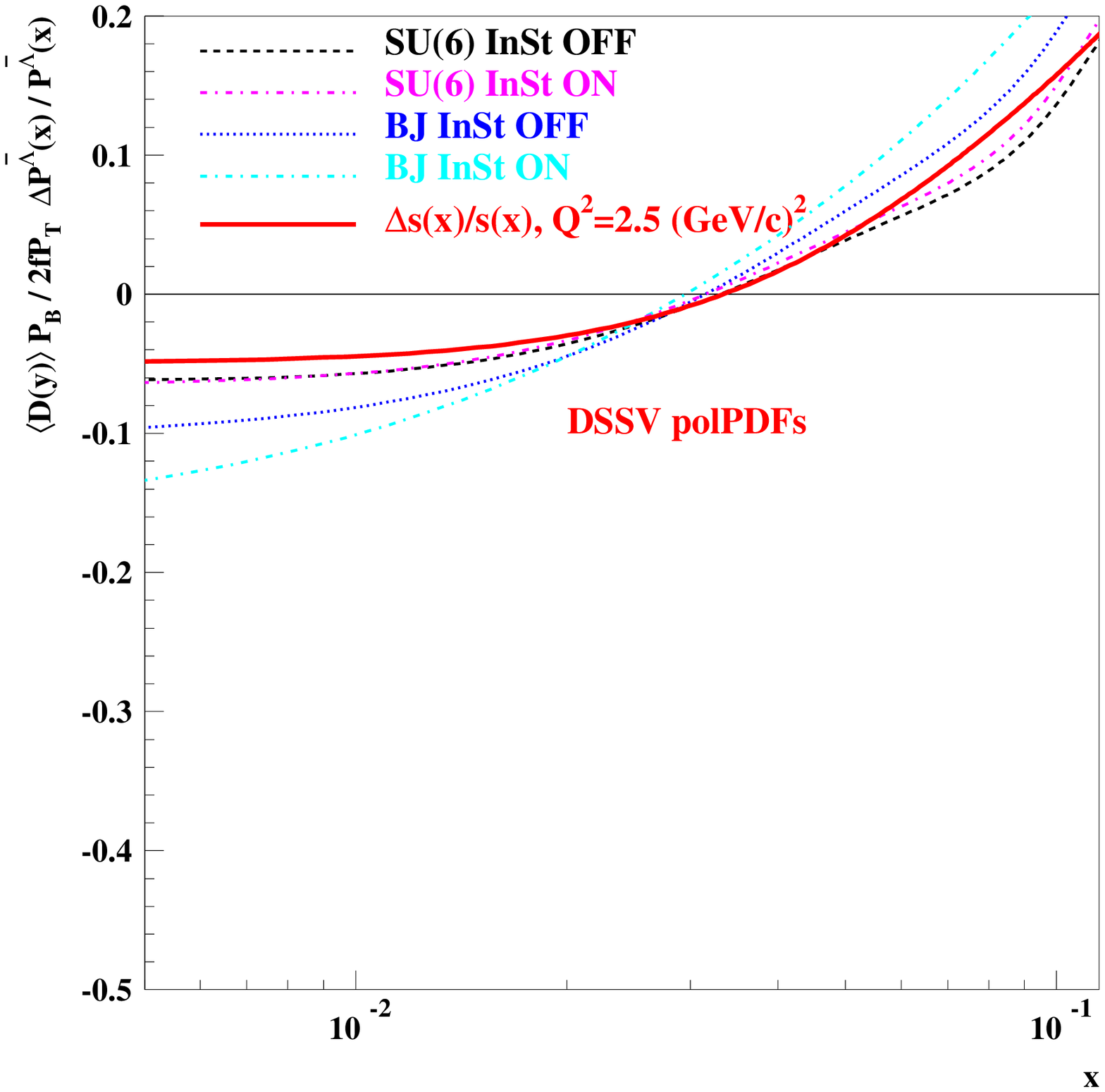, scale=0.4}
    \end{center}
  \end{minipage}
  \hspace{0.6cm}
  \begin{minipage}[t]{.45\textwidth}
    \begin{center}
      \epsfig{file=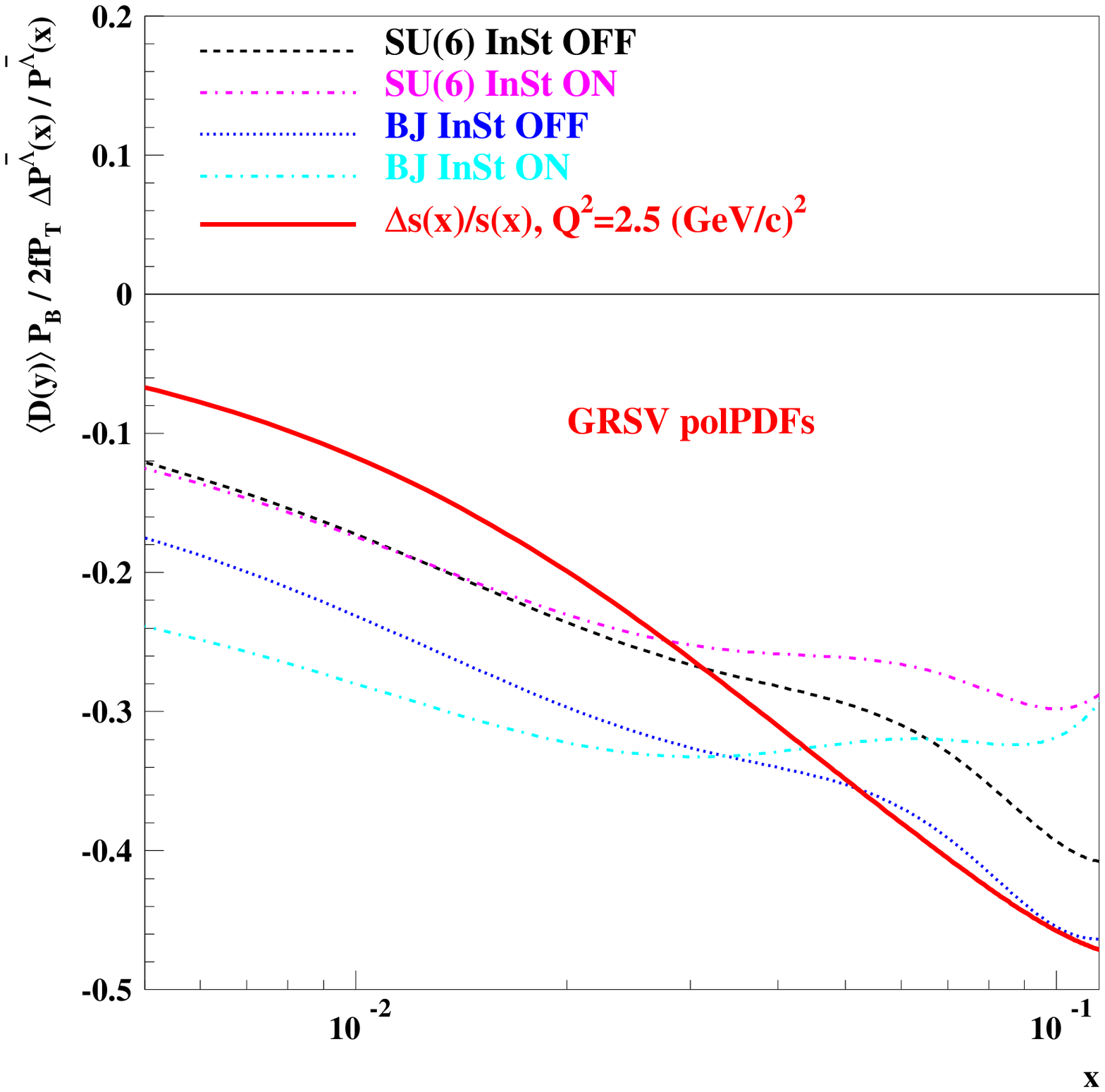, scale=0.4}
    \end{center}
  \end{minipage}
      \caption{Left panel: Polarization asymmetry calculated with DSSV PDFS. Right panel: The same calculated using GRSV PDFs.}
      \label{fig: polasym}
\end{figure}

As for the unpolarized target case we have done the full calculations of the {\it weighted polarization asymmetry} (r.h.s. of Eq. (\ref{ds-s})) using the LEPTO MC with COMPASS cuts. I studied also the influence of polarized intrinsic strangeness when hyperons are produced closer (by rank) to the target remnant than to struck quark. The results of this calculations are presented in Fig. \ref{fig: polasym} where also the polarization of anti-strange quarks (thick continuous line) are presented for for GRSV \cite{grsv} (right panel) and DSSV \cite{dssv} (left panel) polarized PDFs. One can remark that in both cases the contribution from intrinsic strangeness is very small in low ($x<0.02$) $x$ region for SU(6) choice of quark spin transfer (upper dashed and dot-dashed curves). For BJ model the influence of intrinsic strangeness is more visible.

The important conclusion we get from exploiting these different calculations that the weighted polarization asymmetry is closely following the anti-strangeness polarization $\Delta {\bar s} / {\bar s}$. Indeed as one can see in the left panel of Fig. \ref{fig: polasym} the main new feature of the DSSV strangeness PDF, namely, the change of sign of polarized strangeness distribution at $x \simeq 0.03$ is showing up also in all considered model calculations.

\section{Conclusions}

In spite of considerable experimental efforts and repeated performed global fits during the recent years the (polarized) strangeness distribution in nucleon is still not well defined. Therefore every new measurement which can shed light on this subject is very important.

As we have shown the $\bar \Lambda$ polarization measurements in SIDIS of polarized leptons off (un)polarized target can serve as a sensitive anti-strangeness filter. The resent results from COMPASS experiment \cite{Alekseev:2009tv} demonstrate, for example, that CTEQ5L PDFs are better describing the data than GRSV one. The preliminary data on $\bar \Lambda$ polarization dependence on the target polarization were for the first time presented on this workshop \cite{sap}. The statistical accuracy of these data is not so good, but there is a hint on the sign change of $\Delta P^{\bar \Lambda}$.

In conclusion, I would like to stress that new precise data on $\Lambda$ and  ${\bar \Lambda}$ production and polarization in (polarized) SIDIS will represent a great interest to include in global fit for extraction of (un)polarized PDFs and also for deeper understanding of nonperturbative hadronization mechanisms.

\section{Acknowledgments}

It is a pleasure to thank organizers of DIS 09 and especially the convenors of Spin Physics  Carl Gagliardi, Rodolfo Sassot and Gunar Schnell for given to me opportunity to participate at this excellent workshop.




\begin{thebibliography}{99}

\bibitem{url} Slides: \\

\verb$http://indico.cern.ch/contributionDisplay.py?contribId=108&sessionId=4&confId=53294$

\bibitem{ekk}
  J.~R.~Ellis, D.~Kharzeev and A.~Kotzinian,
  Z.\ Phys.\  C {\bf 69}, 467 (1996)
  [arXiv:hep-ph/9506280].

\bibitem{ekn}
  J.~R.~Ellis, A.~Kotzinian and D.~Naumov,
  Eur.\ Phys.\ J.\  C {\bf 25}, 603 (2002)
  [arXiv:hep-ph/0204206]

\bibitem{ekns}
  J.~R.~Ellis, A.~Kotzinian, D.~Naumov and M.~Sapozhnikov,
  Eur.\ Phys.\ J.\  C {\bf 52}, 283 (2007)
  [arXiv:hep-ph/0702222].

\bibitem{sap}
  M.~Sapozhnikov, this proceedings.

\bibitem{Alekseev:2009tv}
  M.~Alekseev  [The COMPASS Collaboration],
  arXiv:0907.0388 [hep-ex].

\bibitem{BJ}
M.~Burkardt, R.~L.~Jaffe, Phys. Rev. Lett. {\bf70} 2537 (1993).

\bibitem{grsv}
  M.~Gluck, E.~Reya, M.~Stratmann and W.~Vogelsang,
  Phys.\ Rev.\  D {\bf 63}, 094005 (2001)
  [arXiv:hep-ph/0011215].

\bibitem{dssv}
  D.~de Florian, R.~Sassot, M.~Stratmann and W.~Vogelsang,
  Phys.\ Rev.\ Lett.\  {\bf 101}, 072001 (2008)
  [arXiv:0804.0422 [hep-ph]].



\end{thebibliography}
\end{document}